\documentclass[useAMS,usenatbib]{mn2e}
\usepackage{graphics}
\usepackage{epsfig}

\title[Optical and Infrared Light Curves of V395 Car]{Optical and Infrared Light Curves of the Eclipsing X-ray Binary
V395 Car = 2S 0921--630}

\author[T. A. Ashcraft et al.]{T. A. Ashcraft$^{1}$, R. I. Hynes$^{2}$, and E. L. Robinson$^{3}$\\
$^{1}$School of Earth and Space Exploration, Arizona State University, Tempe, AZ 85287-1404, USA\\
$^{2}$Department of Physics and Astronomy, Louisiana State
University, Baton Rouge, Louisiana 70803, USA\\
$^{3}$Department of Astronomy, The University of Texas at Austin, 
1 University Station C1400, Austin, Texas 78712, USA}
\begin{document}

\date{In press}

\pagerange{\pageref{firstpage}--\pageref{lastpage}} \pubyear{2012}

\maketitle

\label{firstpage}

\begin{abstract}
  We present results of optical and infrared photometric monitoring of
  the eclipsing low-mass X-ray binary V395 Car (2S~0921--630).  Our
  observations reveal a clear, repeating orbital modulation with an
  amplitude of about one magnitude in $B$, and $V$ and a little less
  in $J$. Combining our data with archival observations spanning about
  20\,years, we derive an updated ephemeris with orbital period
  $9.0026\pm0.0001$\,d.  We attribute the modulation to a combination
  of the changing aspect of the irradiated face of the companion star
  and eclipses of the accretion disk around the neutron star. Both
  appear to be necessary as a secondary eclipse of the companion star
  is clearly seen. We model the $B$, $V$, and $J$ lightcurves using a
  simple model of an accretion disk and companion star and find a good
  fit is possible for binary inclinations of $82.2\pm1.0^{\circ}$.  We
  estimate the irradiating luminosity to be about
  $8\times10^{35}$\,erg\,s$^{-1}$, in good agreement with X-ray constraints.
\end{abstract}


\section{Introduction}

2S~0921--630 was discovered as an X-ray source by \citet{Li:1978a} using
SAS-3 and was identified with an approximate 17th magnitude star, V395~Car. The
system shows partial eclipses in both the optical and X-ray bands
\citep{Branduardi-Raymont:1983a, Chevalier:1982a, Mason:1987a}. The
orbital period is about 9\,days and the system must have an
inclination between $70^{\circ}$ and $90^{\circ}$. Optical dips of up
to 2 magnitudes deep \citep{Krzeminski:1991a} have been present in
optical light curves of V395~Car, but no complete orbital
lightcurve has yet been published. The companion star has been
identified as a K0\,{\sc iii} star \citep{Shahbaz:1999a}.

Several attempts have been made to constrain the system parameters of
V395~Car using the photospheric absorption lines from the companion
star, which are detectable in spite of the large contamination by disk
flux
\citep{Shahbaz:1999a,Shahbaz:2004a,Jonker:2005a,Shahbaz:2007a,Steeghs:2007a}.
From measurements of the radial velocity curve and rotational
broadening, and assuming an appropriate inclination for an eclipsing
system, it is possible to solve for the masses of both objects. The
remaining uncertainty is in the `K-correction', which accounts for
suppression of absorption lines on the inner face of the companion
star by irradiation. Assuming negligible K-correction and an
inclination of $i=83^{\circ}$, \citet{Steeghs:2007a} deduce a compact
object mass of $M_1=1.44\pm0.1$\,M$_{\odot}$, a companion star mass of
$M_2=0.35\pm0.03$\,M$_{\odot}$, and hence a mass ratio of
$q=0.24\pm0.02$. \citet{Shahbaz:2007a} assume $i=75^{\circ}$ and
deduce $M_1 = 1.37\pm0.13$\,M$_{\odot}$ and $q=0.281\pm0.034$. We note
that the assumed inclination does not affect the derived mass ratio,
and that the two estimates of $q$ are consistent. The mass estimates,
too, agree, and both are consistent with a canonical mass neutron
star, in contrast to earlier mass estimates
\citep{Shahbaz:2004a,Jonker:2005a}, which used the overestimated
rotational broadening measurement of \citet{Shahbaz:1999a}.

Because V395~Car eclipses, the inclination is
relatively well constrained and the derived parameters are relatively
insensitive to the remaining inclination uncertainty. Nonetheless, it
is still of interest to examine the orbital lightcurve of the binary,
and in fact it is an advantage that the system parameters are already
well constrained, reducing the uncertainty in the interpretation of
the lightcurves.

In this paper we present optical and IR photometry from two observing
seasons totalling 187 nights of monitoring of V395~Car obtained to
investigate both the orbital and longer-term lightcurve. We begin by
describing our data in Section~\ref{DataSection}.
Section~\ref{LongtermSection} presents the long-term lightcurve, and
Section~\ref{PeriodSection} proceeds to derive an improved measurement
of the orbital period and ephemeris zero-point based on these data
together with archival measurements. With the period firmly
established, we present orbital lightcurves in the three bands in
Section~\ref{OrbitalSection} and model them in
Section~\ref{ModelingSection}. We discuss the implications of our
results in Section~\ref{DiscussionSection} and summarise our
conclusions in Section~\ref{ConclusionSection}.

\section{Observations and Data Reduction}
\label{DataSection}

\subsection{Optical and Infrared Data}

Using the SMARTS 1.3\,m telescope at Cerro Tololo Inter-American
Observatory with the dual-channel optical/IR ANDICAM, images were
taken almost daily from 2004 November through 2005 July and in 2005
December of V395~Car in the $B$, $V$, and $J$ bands.  For various
reasons such as equipment and weather problems, there are a few long
gaps in the data. The overall quality of the images was
typically good, with seeing of 1--2\,arcsec.  Usable data were obtained
on a total of 187 nights.

\begin{figure}
\epsfig{width=3.3in,file=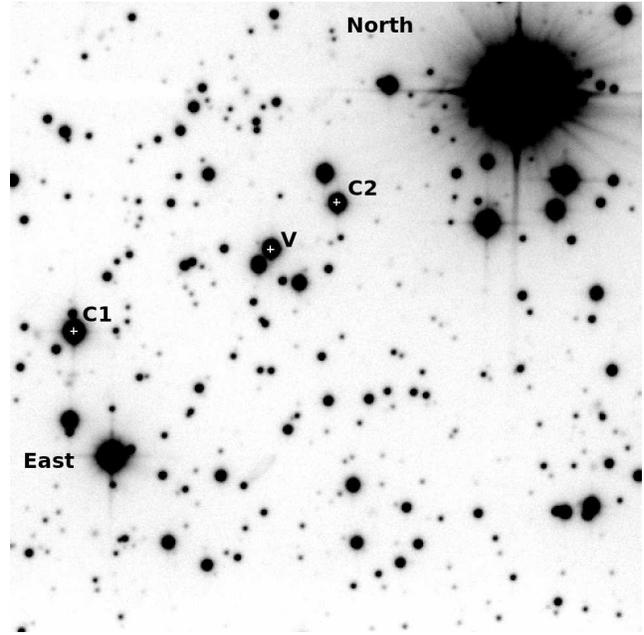}
\caption{Finder chart based on the average of all of our $V$ band
  images.  V395~Car is labeled V, the comparison star used for optical
images is C1 and the IR comparison is C2.  The field of view is
3\,arcmin square.}
\label{FinderFig}
\end{figure}

On each night two images were taken in the $V$ band with 130\,s
exposure time and two in the $B$ band with a 100\,s exposure
time. Pipeline data reduction for the optical frames was done prior to
receiving the images and appeared satisfactory, with no residual
artifacts of poor bias corretion or flat-fielding. We show the central
portion of the average $V$ frame in Figure~\ref{FinderFig}.  Each
night differential aperture photometry was performed relative to a
bright reference star for the target and three other comparison stars
used to check the stability using standard IRAF\footnote{IRAF is
  distributed by the National Optical Astronomy Observatories, which
  are operated by the Association of Universities for Research in
  Astronomy, Inc., under cooperative agreement with the National
  Science Foundation.}  routines.

Additionally, nine images with 50\,s exposure time were taken each
night in the $J$ band, with each image shifted with respect to the
others. The nine images were median combined in IRAF to create a sky
image which was subtracted from each individual image before
flat-fielding using dome flats. The reduced images were shifted and
averaged to produce one final image per night.

\section{Long-term Lightcurves}
\label{LongtermSection}

\begin{figure}
\epsfig{width=3.5in,file=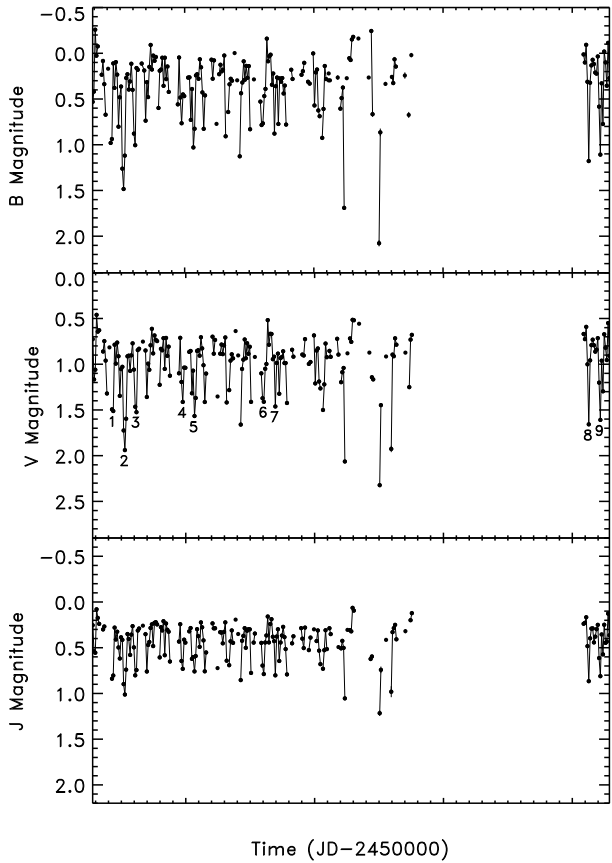}
\caption{Plot of the long-term lightcurves in the B, V, and J
  bands. For the B and V bands the data points for the two images
  taken each night were averaged together for clarity. The dates of
  the 9 eclipses used to find the ephemeris are labelled 1-9.}
\label{LongtermFig}
\end{figure}

To obtain the lightcurves we performed differential photometry
relative to a comparison star. The same comparison star was used for
the $B$ and $V$ bands (C1 in Figure~\ref{FInderFig}) but a different
one was necessary for $J$, C2, as C1 was not consistently located in
the smaller IR field of view.  We show the combined long-term
lightcurve in Figure~\ref{LongtermFig}.  In this, and all plots in
this paper, we show differential magnitudes relative to the comparison
star used. The lightcurves in all the bands exhibit only the basic
eclipsing binary modulation. There are no indications of any longterm
variation in the out of eclipse lightcurve.

\section{Period Analysis}
\label{PeriodSection}

\begin{figure}
\epsfig{angle=90,width=3.4in,file=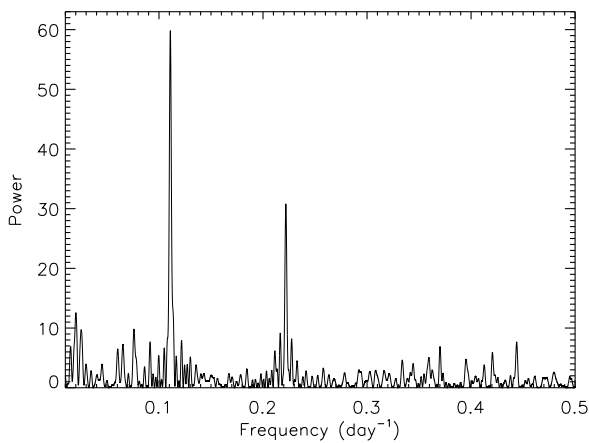}
\caption{Periodogram of the V band. It only shows frequencies up to
  0.5 day$^{-1}$ since higher ones are duplicates of their lower frequency
  counterpart. The peak of 9.01632 days clearly dominates over the
  peak caused by the secondary eclipse.}
\label{ScargleFig}
\end{figure}

For each band, $B$, $V$, and $J$, we computed the periodogram using
the Lomb-Scargle technique to investigate the different possible
periods. We show the $V$ band periodogram in Figure~\ref{ScargleFig}.
Two strong peaks were seen, a strong fundamental near 9\,days and a
first harmonic. For the $B$ and $V$ bands peak at 9\,days is
dominant. The $J$ band is more affected by the secondary eclipse,
which caused the first harmonic to slightly dominate. We measured the
fundamental periods to be 9.01307\,days ($B$), 9.01632\,days ($V$),
and 9.02283\,days ($J$).  A period of 9.017 $\pm$ 0.005 days came from
averaging together the period from each filter. The accuracy of the
period found from this method is not high because it is sensitive to
slight changes in the data. Furthermore, since all of the filters were
obtained at approximately the same time, they are not fully
independent, and the uncertainty quoted is an underestimate.  The mean
period does agree with the periods others have previously found:
$9.0035\pm0.0029$\,days \citep{Shahbaz:2004a} and
$9.006\pm0.007$\,days \citep{Jonker:2005a}.

We can obtain a much more precise measurement of the period by
combining our data with previously published data.  We measured the time of
nine different mid-eclipses for which there were data available for
three consecutive days around the time of mid-eclipse. Averaging these
nine eclipses together, an ephemeris of eclipse of $T_0 = 2453397.52
\pm0.08$ in HJD was determined.  The uncertainty on this was large and
we found that when we folded and fit the lightcurve in
Section~\ref{ModelingSection} a phase offset of $-0.0271\pm0.0024$ was
required.  Correcting for this offset, we refine our eclipse ephemeris
to $T_0=2453397.28\pm0.02$.  Using the previously calculated time of
mid-eclipse of 2446249.18 from the ephemeris of \citet{Mason:1987a}
and the newly calculated one, and assuming that the
\citet{Shahbaz:2004a} period (the most precise previously available)
is accurate, a refined orbital period could be 8.9913 $\pm$ 0.0001
days (795 cycles), 9.0026 $\pm$ 0.0001 days (794 cycles), or 9.0140
$\pm$ 0.0001 days (793 cycles), where only 9.0026\,days is consistent
within the uncertainties of Shahbaz at the $3\sigma$ level.

\begin{figure}
\epsfig{angle=90,width=3.4in,file=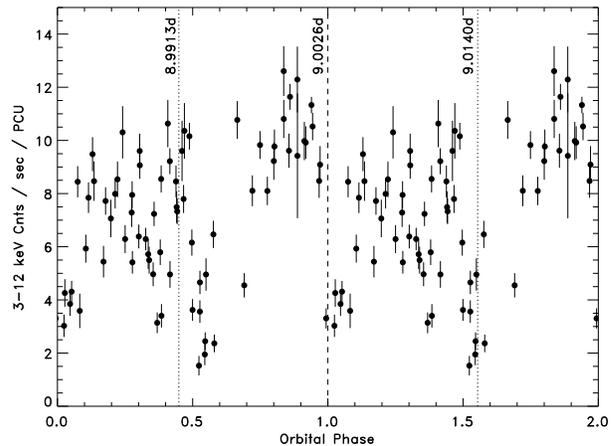}
\caption{{\it RXTE}/PCA lightcurve from 1996 folded on the 9.0026\,d
  period.  The dashed lines indicate times of predicted eclipse for
  not only this period, but for the other two candidates.  Since the
  X-ray observations were concentrated in a short period, the primary
  effect of different assumed orbital periods is to shift the minimum
  as indicated.}
\label{XRayFig}
\end{figure}

To attempt to confirm the identification of the correct alias, we also
examined 75 archival {\it Rossi X-ray Timing Explorer} PCA
observations obtained from April 12 to May 2, 1996.  We extracted
3--12\,keV background subtracted fluxes from Standard 2 data and
combined each ($\sim1000$\,s) exposure into a single average point.
We show these data folded on the 9.0026\,d period in
Figure~\ref{XRayFig}; folding on the alternative periods results in
lightcurves that are just shifted, so rather than replicate the data
we show where phase zero is predicted to fall for each of these
periods.

2S~0921--630 is known to show X-ray eclipses at phase zero
\citep{Branduardi-Raymont:1983a}.  From the X-ray lightcurve we can
therefore rule out the period of 8.9913 days since no structure is
seen at phase 0. The 9.0026\,d period does produce an apparent eclipse
centred at phase zero as expected.  However the 9.0140\,d period also
produces a minimum at around the same phase, albeit a less uniform
one.  The X-ray eclipse presented by \citet{Mason:1987a} has a full
width of $\sim 0.15$ in orbital phase, and drops to about 25\,percent
of the peak intensity.  This is a rather good match to the eclipse
width and depth we obtain from a 9.0026\,day period supporting that
identification.  If this is correct, then there is a broad and
irregular dipping structure from phase 0.5--0.7 which would be
consistent with X-ray dipping seen in other high inclination LMXBs.
While we cannot absolutely rule out a 9.0140\,d period, we conclude
that 9.0026\,d is most likely to be the true orbital period
(consistent with the measurement of \citealt{Shahbaz:2004a}), and so
derive an updated ephemeris:
$$T_0 ({\rm HJD}) = 2453397.28(2) + 9.0026(1)E$$
where $T_0$ is the time of ecclipse.  We note that the difference
between the 9.0026\,day and 9.0140\,day period has negligible effect
on the folding of our lightcurves and so does not affect the
subsequent analysis or modeling.

\section{Orbital Lightcurves} 
\label{OrbitalSection}

\begin{figure}
\epsfig{width=3.5in,file=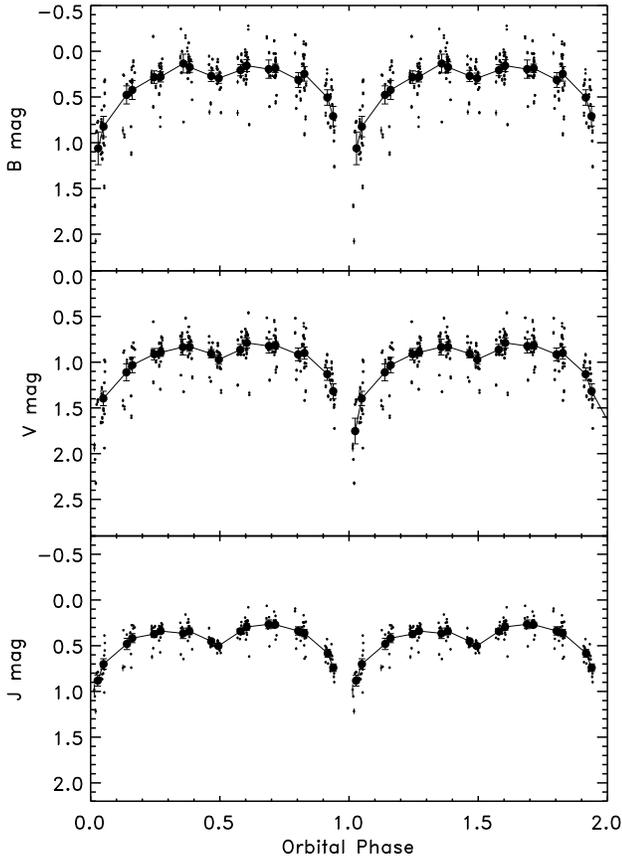}
\caption{Orbital lightcurves of V395 Car in all three bands $B$, $V$, and
  $J$ created by folding the data over a period of 9.0026 days. Larger
  points are phase-binned averages and are joined by lines for
  clarity.}
\label{FoldFig}
\end{figure}

Fig.~\ref{FoldFig} shows the orbital lightcurve based on our
ephemeris. Since the period is almost a whole number of days the
points are bunched up in phase leading to incomplete coverage,
especially during primary eclipse.

Each band clearly shows a primary deep eclipse and shallower secondary
eclipse with the primary eclipse having the greatest depth in the $B$
band and the secondary being the most prominent in the infrared $J$
band. The depths of the primary eclipse are $\sim2.5$\,mag in $B$,
$\sim1.9$\,mag in $V$, and $\sim1.1$\,mag in $J$.

\section{Modelling Lightcurves}
\label{ModelingSection}

We modelled the lightcurves using the XRbinary\footnote{A full
  description of the program is available at
  http://pisces.as.utexas.edu/robinson/XRbinary.pdf.} code, version
2.1.  This code has previously been used for the LMXBs UW~CrB
\citep{Mason:2008a}, 4U~1822--371 \citep{Bayless:2010a}, and
4U~1957+115 \citep{Bayless:2011a}, as well as
for the cataclysmic variable SS~Cyg \citep{Bitner:2007a}.  The general
approach is to model the disk and companion star as a series of tiles.
Irradiation is calculated by ray-tracing, allowing for rather complex
disk geometries.  For this analysis, we have used only the simplest
symmetric disk model, which nonetheless provides quite a good fit to the
multi-band data.

The system parameters of V395~Car are relatively well known, at least
compared to many neutron star LMXBs.  The mass ratio, independent of
inclination, has been independently estimated at $q=0.281\pm0.034$
\citep{Shahbaz:2007a}, and $q=0.24\pm0.02$ \citep{Steeghs:2007a}.  Both
groups used essentially the same method, measuring the radial velocity
semi-amplitude and rotational broadening of the companion star and
deducing the mass ratio using the relationship of \citet{Wade:1988a}.
We adopt $q=0.26$ for this analysis as an average of the two
measurements.  \citet{Steeghs:2007a} noted that for their mass ratio,
the X-ray eclipse observation of \citet{Mason:1987a} implies an
inclination about $i=83^{\circ}$, but we will leave the inclination as
a free parameter when modelling the lightcurves.

We assume illumination of the disk and companion star by a source of
X-rays at a height above the accretion disk intended
to mimic an accretion disk corona (ADC).  The X-ray source is
specified by its luminosity and the height above and below the plane
(it is treated as a lamppost).  In the limit of placing this at a
negligible height above the accretion disk, this also would represent
central illumination from the neutron star.  We also experimented with
models in which the accretion disk was simply given a parameterised
temperature distribution, $T\propto R^{-0.5}$ or $T\propto R^{-0.75}$,
but found these produced inferior fits compared to the lamppost ADC
model.

For the disk geometry, we assume a thin, flared, axisymmetric disk,
with $H/R \propto R^{1.21}$, where $H$ is the local disk height and
$R$ is the radius. The height and radius of the outer edge of the disk
are free parameters. We adopted an exponent of 1.21 as representative
of values proposed in the literature rather than strongly motivated in
its own right, but we also repeated the analysis with $H/R \propto
R^{1.10}$ and $H/R \propto R^{1.29}$ and found negligible difference,
and in particular no effect on the preferred inclination. This is
expected as our fits preferred a very thin disk and a significantly
elevated X-ray source. With these parameters the shape of the disk has
relatively little effect and irradiation is possible even if the real
disk profile is convex \citep{Dubus:1999a}. The outer radius of the
disk is a free parameter but since the outer radius is limited by
tidal truncation, we constrain it to be less than 0.9\,$R_{\rm lobe}$
\citep{Whitehurst:1991a}.

Reprocessing of X-rays will also be affected by a local albedo. We
assume all X-rays incident on the companion are reprocessed, but that
a fraction of those incident on the disk might be Compton reflected,
with the fraction left as a free parameter. While some Compton
reflection from the companion star is also possible, we fold this
uncertainty into the X-ray luminosity, as there are really only two
independent parameters that can be constrained here: the heating of
the disk and the heating of the donor star. The assumption being made
is that the accretion disk will be more highly ionised than the
companion star, and irradiated predominantly at a steeper angle of
incidence, so would be expected to thermally reprocess a smaller
fraction of X-rays than the donor star.

To summarise this discussion, the free parameters of the model are the
binary inclination ($i$), the disk outer radius ($R_{\rm out}$) and
height ($H_{\rm out}$), the X-ray luminosity ($L_{\rm X}$),
the height of the X-ray source above the plane ($H_{\rm ADC}$), and the relative
albedo of the disk and donor star.  We performed a series of fits with
fixed trial inclinations, as this is the parameter which we care most
about, while leaving other parameters free to vary (subject to
constraints on the disk radius and relative albedo).  All fits were
begun using a downhill simplex (amoeba) algorithm.  Multiple
starting simplexes were tested, and each fit was refined using
Powell's method.  

\begin{figure}
\epsfig{width=3.5in,file=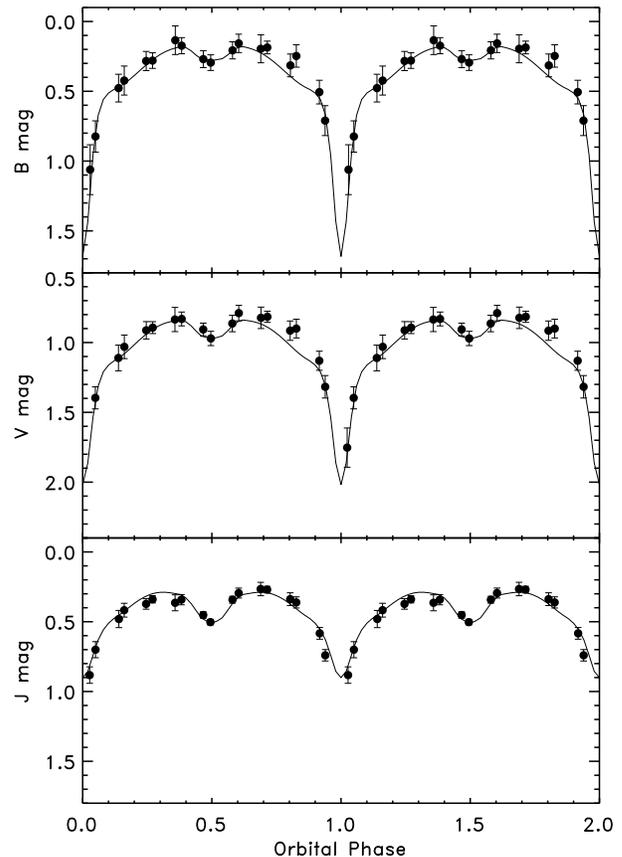}
\caption{The orbital binned lightcurve fitted with the best-fitting solution.}
\label{ModelFig}
\end{figure}

We fitted models at $0.1^{\circ}$ intervals from 80--85$^{\circ}$, and
more coarsely beyond that. We find a best fitting value of
$i=82.2\pm1.0^{\circ}$ (1$\sigma$). We also ran fits for $q=0.24$ and
$q=0.28$.  These were minimally different, with the best fitting
inclination shifted by no more than 0.2$^{\circ}$, a difference
negligible compared to our statistical uncertainty.  The best-fitting
model is shown in Fig.~\ref{ModelFig}. It corresponds to
$i=82.2^{\circ}$, $L_{\rm X}=8\times10^{35}$\,erg\,s$^{-1}$, $R_{\rm
  disk}=0.89R_{\rm lobe}$, $H/R=0.001$, $H_{ADC}=0.16a$, and equal
disk and donor star albedos.  It is apparent that the deepest points
of the model primary eclipse are much deeper than the binned points,
especially at shorter wavelengths.  This is a consequence of the
sampling however, and examining Fig.~\ref{FoldFig}, we can see that
the same trend is present in the unbinned data.

\section{Discussion}
\label{DiscussionSection}

The inclination we derive is consistent with previous estimates.  For
example, \citet{Steeghs:2007a} comment that for their mass ratio, an
inclination of 83$^{\circ}$ is implied by the X-ray lightcurve of
\citet{Mason:1987a}.  Consequently, our inclination does not change
current estimates of the neutron star mass significantly, especially
as the derived mass is relatively insensitive to inclination for high
inclinations.  Nonetheless, pinning down the inclination does increase
the confidence in this measurement.  \citet{Steeghs:2007a} estimated
$M_1 = 1.44\pm0.10$\,M$_{\odot}$ by assuming $i=83.0$ with no
uncertainty in inclination.  Adopting instead $i=82.2\pm1.0^{\circ}$
and performing a similar Monte-Carlo simulation to that used by the
authors does not change either the derived value or its uncertainty.

While our estimate of the luminosity of the X-ray source is indirect,
it is consistent with (and independent of) that of
\citet{Kallman:2003a}; $L_{\rm X} < 10^{36}$\,erg\,s$^{-1}$ for cases
in which we see the full X-ray luminosity.  The luminosity sensitivity
of our lightcurve fitting arises because the X-ray luminosity sets the
temperatures of the accretion disk and companion star.  Different
temperatures will change the relative depths of primary and secondary
eclipses as a function of bandpass, so the broad wavelength
coverage of our data is what leads to our sensitivity to the X-ray
luminosity.  Our measurement is very model dependent and cannot be a
precise one, but is consistent with the model of
\citet{Kallman:2003a} in which we see the full X-ray luminosity
rather than a small fraction of it as in canonical ADC sources.

Of the other parameters, the disk radius implies a tidally truncated
disk, as might be expected.  The disk thickness is smaller than would
typically be expected, but may simply indicate that the disk thickness
is negligible compared to the vertical extent of the X-ray emitting
region, which we infer to be rather large.  A large X-ray emitting
region is also required by the presence of partial eclipses at X-ray
wavelengths, so is not surprising.

Finally, we note that there is a small residual asymmetry in the
lightcurves.  This is common in LMXBs and usually attributed to
non-axisymmetric structure in the accretion disk.  Given that it is a
small effect and that our lightcurves are not well sampled we do not
try to model the asymmetry numerically.

\section{Conclusions}
\label{ConclusionSection}

We have obtained the first full, and multi-colour, orbital lightcurve
of the long-period LMXB V395~Car. We refine the orbital period to
$9.0026\pm0.0001$\,days and present an updated time of minimum, $T_0 =
2453397.28\pm0.02$ (HJD). We successfully fit our $BVJ$ lightcurves
using the XRbinary code and derive a binary inclination of
$i=82.2\pm1.0^{\circ}$. This is consistent with previous constraints
and does not change previously published compact object mass
estimates.

We also estimate an X-ray luminosity of $10^{36}$\,erg\,s$^{-1}$,
consistent with X-ray estimates of \citet{Kallman:2003a}. This makes
V395~Car a relatively low mass-transfer rate system in sharp contrast
to 4U~1822--371 where an accretion rate close to the Eddington limit
is inferred. \citep{Bayless:2010a}.

With the period so close to an integer days, one of the major
limitations of our dataset remains the incomplete sampling in orbital
phase.  Further progress will require a completely sampled lightcurve,
and consequently a multi-site campaign spanning a range of longitudes.

\section*{Acknowledgments}

This work uses data obtained on the 1.3\,m telescope at Cerro Tololo
Inter-American Observatory which is operated by the SMARTS Consortium.
This research has made use of data obtained through the High Energy
Astrophysics Science Archive Research Center Online Service, provided
by the NASA/Goddard Space Flight Center and has also made use of the
NASA ADS Abstract Service.  R.I.H.  acknowledges support from
NASA/Louisiana Board of Regents grant NNX07AT62A/LEQSF(2007-10)
Phase3-02 and National Science Foundation Grant No.\ AST-0908789.

\label{lastpage}

\end{document}